\documentclass[preprint,preprintnumbers,amsmath,amssymb,APS]{revtex4}
\usepackage{amsmath}
\usepackage{graphicx}
\usepackage{dcolumn}
\usepackage{array}
\usepackage{bm}
\usepackage{subfigure}

\usepackage[USenglish]{babel} 
\usepackage{graphicx}
\usepackage{dcolumn}
\usepackage{epstopdf}
\usepackage{bm}
\usepackage{color} 
\usepackage{comment} 
\usepackage{ulem} 
\usepackage{verbatim}

\begin{document}


\title{Realization of One-Way Electromagnetic Modes at the Interface Between Two Dissimilar Metals}

\author{Mehul Dixit$^{1,2}$ and David Stroud$^1$}

\affiliation{$^1$Department of Physics, The Ohio State University,
Columbus, OH 43210}

\affiliation{$^2$ Department of Physics, University of Missouri, Columbia, MO 65211}

\date{\today}

\begin{abstract}

We calculate the dispersion relations for electromagnetic waves propagating at the interface between two dissimilar Drude metals in an external magnetic field
${\bf B}$  parallel to the interface.  The propagating modes are bound to the inteface and travel perpendicular to ${\bf B}$.   In certain frequency ranges, the waves can propagate
in one direction only.   The frequency range for these one-way modes increases with increasing ${\bf B}$.    One group of modes occurs at moderate frequencies, between the lower and upper plasma frequencies of the
two metals.  The other occurs at much lower frequencies,
between their lower and upper cyclotron frequencies. We discuss possible ways to realize such modes
in real materials, including dissimilar superconductors.

\end{abstract}


\maketitle

\newpage


Electromagnetic waveguides are important  in optical integrated circuits, among other applications\cite{Krauss03}. Electromagnetic wave propagation in such waveguides is affected by the presence of disorder, which can cause back-scattering, leading to losses.  Controlling such scattering is especially important for nanodevices\cite{Yu08} and for slow light systems, which are of current interest for optical signal processing applications\cite{Povinelli04}. 
One-way electromagnetic waveguides are of particular interest in this regard.  
In such waveguides, there is a frequency range where waves propagate in one direction only, and cannot be scattered into the reverse direction.  

The earliest proposals for the realization of electromagnetic one-way waveguides were presented in Refs.\ \cite{RaghuHaldane08}, \cite{Yu08}, and \cite{Wang08}.   Ref.\ \cite{RaghuHaldane08} showed the existence of one-way electromagnetic modes at the interface between two special classes of two-dimensional (2D), Faraday-active photonic crystals with
band gaps derived from a Dirac point in the photonic band structure.  Ref.\ \cite{Wang08} showed that a Dirac point derived band gap is not necessary.  In particular, they found that some
transverse magnetic (TM) modes in a 2D square lattice of yttrium iron garnet (YIG) rods were one-way.

Ref.\ \cite{Yu08} showed that one-way electromagnetic modes can exist at the interface of a plasmonic metal,
which was approximated as lossless,
 and a dielectric photonic crystal in the presence of a static magnetic field parallel to the interface.  They found that surface waves propagating parallel to the 
interface but perpendicular to the magnetic
field are one-way in certain frequency regimes, provided that the surface plasmon frequency of the metal lies within the band-gap of the photonic crystal. The surface plasmon dispersion in such a waveguide structure was also shown to be nonreciprocal, i.~e. having a dispersion relation such that  $\omega(\mathbf{k}) \neq \omega(-\mathbf{k})$, with
one-way propagation at certain frequencies within the band-gap of the photonic crystal.

In the present work, we extend the model of Ref.\ \cite{Yu08} to electromagnetic surface plasmon modes bound to the interface of two dissimilar metals in an external magnetic field parallel to the interface.  
Although, forr computational convenience, we solve for the dispersion relations in the absence of  loss, we believe that these relations would be similar in
the more realistic case of nonzero, but weak, damping.
 As in the model of Ref.\ \cite{Yu08}, we find that there can exist frequency regimes which support one-way electromagnetic wave propagation.    
There prove to be two distinct regimes of apparently one-way wave propagation.  
 One is at moderate frequencies, between the lower and upper plasma frequencies of the two metals.  The other range is at much lower frequencies, between the lower and upper cyclotron frequencies of the metals.  

\begin{figure}[tbh]
\begin{center}
\includegraphics[width=\columnwidth]{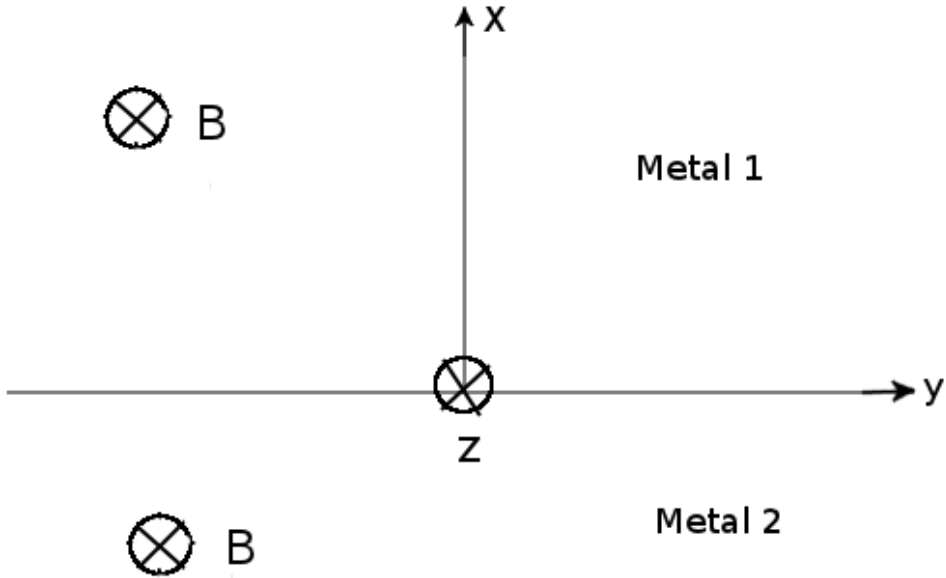}
\caption{Schematic of the waveguide configuration described in the text.  Metals 1 and 2 have plasma frequencies
$\omega_{p1}$ and $\omega_{p2}$, and occupy the half-spaces $x>0$ and
$x<0$, respectively.  A static magnetic field ${\bf B} = B\hat{z}$ is applied along the z axis.  
Surface plasmon waves propagating along the y axis show one-way behavior in certain frequency ranges.}
\label{waveguideconfiguration}
\end{center}
\end{figure}

The structure we use consists of two dissimilar semi-infinite non-magnetic metals (relative permeability $\mu_{r} = 1$ for both), with a planar interface which we take to lie in the yz plane.
The two metals are assumed to have bulk plasma frequencies $\omega_{p1}$ and $\omega_{p2}\neq\omega_{p1}$.   There is assumed to be a static external magnetic field ${\bf B} = B\hat{z}$ parallel to the interface along the z direction (See Fig.~\ref{waveguideconfiguration}). The external magnetic field breaks time-reversal symmetry and induces anisotropy in the dielectric tensors $\epsilon_{i}\left(\omega\right)$ ($i = 1,2$) of the two metals.   The dielectric tensors of the two metals can be obtained from the classical
equation of motion for an electron in a magnetic field\cite{Jackson1}.  For an electron in the i$^{th}$ metal (i = 1, 2) of momentum $\mathbf{p}_i$, effective mass $m_i^*$, charge $-e$,
and damping time $\tau_i$ in a magnetic field $\mathbf{B}$ and electric field $\mathbf{E}$, these take the form (we use SI units throughout)
$\frac{d\mathbf{p}_i}{dt} = -\frac{\mathbf{p}_i}{\tau_i} - e\frac{1}{m^{*}_i}\left[\mathbf{E}+\mathbf{p}_i\times \mathbf{B}\right]$.
These can be solved in the standard way, assuming an electric field of frequency $\omega$, to obtain the complex dielectric tensor $\epsilon_i(\omega)$ of the i$^{th}$ metal:
\begin{align}
\label{epsilon1}
\epsilon_i\left(\omega\right) = 1 - \frac{\omega^2_{pi}}{\left(\omega + \frac{i}{\tau_i}\right)^2 - \omega^2_{Bi}}\times \nonumber \\
\left(\begin{array}{c c c}
	\left(1 + \frac{i}{\omega\tau_i}\right) & -i\frac{\omega_{Bi}}{\omega} & 0 \\
	i\frac{\omega_{Bi}}{\omega} & \left(1 + \frac{i}{\omega\tau_i}\right) & 0\\
	0 & 0 & \frac{\left(\omega + \frac{i}{\tau_i}\right)^2 - \omega^2_{Bi}}{\omega\left(\omega + \frac{i}{\tau}\right)} \end{array}\right),
\end{align}
where $\omega_{Bi} = eB/m_i^*$.
In the limit $\tau_i \rightarrow \infty$, corresponding to zero losses, Eq.\ (\ref{epsilon1}) reduces to
\begin{align}
\label{epsilon}
\epsilon_i\left(\omega\right) = 1 - \frac{\omega^2_{pi}}{\omega^2 - \omega^2_{Bi}}\left(\begin{array}{c c c}
	1  & -i\frac{\omega_{Bi}}{\omega} & 0 \\
	i\frac{\omega_{Bi}}{\omega} & 1 & 0\\
	0 & 0 & \frac{\omega^2 - \omega^2_{Bi}}{\omega^{2}} \end{array}\right).
\end{align}

We wish to find solutions to Maxwell's equations corresponding to electromagnetic surface plasmon modes bound to the planar interface and propagating in the $y$ direction, normal
to the external magnetic field ${\bf B} = B\hat{z}$ (see Fig.\ 1).  The half-space $x > 0$ is assumed occupied by material 1; and that in $x<0$, by material 2.    Thus we seek modes
with electric fields
\begin{eqnarray}
\label{elecfield}
\mathbf{E}\left(\mathbf{r},t\right) \equiv \left\{ \begin{array}{c c} \mathbf{E}_{1} = \left({E}_{1x}\hat{x} + {E}_{1y}\hat{y}\right)\, e^{i k_{y}y - k_{1x}x}e^{-i \omega t}; \quad x \geq 0,\\
\mathbf{E}_{2} = \left({E}_{2x}\hat{x} + {E}_{2y}\hat{y}\right)\, e^{i k_{y}y + k_{2x}x}e^{-i \omega t}; \quad x < 0,
\end{array} \right.
\end{eqnarray}
with $k_{1x}, k_{2x} > 0$  for surface waves. The corresponding magnetic field ${\bf B}\left(\mathbf{r},t\right)$ is 
\begin{eqnarray}
\label{magfield}
\mathbf{B}\left(\mathbf{r},t\right) \equiv \left\{ \begin{array}{c c} \mathbf{B}_{1} = {B}_{1}\hat{z}\, e^{i k_{y}y - k_{1x}x}e^{-i \omega t}; \quad x \geq 0,\\
\mathbf{B}_{2} = {B}_{2}\hat{z}\, e^{i k_{y}y + k_{2x}x}e^{-i \omega t}; \quad x < 0.
\end{array} \right.
\end{eqnarray}
The magnetic field amplitude $B_{i}$ ($i = 1,2 $) is obtained from Faraday's Law, expressed for fields of frequency $\omega$ as 
$i\omega\mathbf{B}_{i} = \nabla \times \mathbf{E}_{i}$, which leads to the following relation between $B_i$ and $E_i$:
\begin{align}
{B}_{i} = \frac{i}{\omega} \left(\pm k_{ix} E_{iy} + i E_{ix} k_{y}\right),
\end{align}
where the upper sign is for $i = 1$ and the lower for $i = 2$.

Eliminating the magnetic field using the two Maxwell curl equations gives the following relation between ${\bf E}_i$ and ${\bf k}_i$:
\begin{equation}
\label{maxwelleqninmedium}
k^2_{i}\mathbf{E}_{i} - \mathbf{k}_{i}\left(\mathbf{k}_{i}\cdotp\mathbf{E}_{i}\right) = \frac{\omega^2}{c^2} \epsilon_{i}\left(\omega\right) \mathbf{E}_{i},
\end{equation}
where 
\begin{equation}
\label{eqnki}
\mathbf{k}_{i} = \left(\pm i k_{ix} \hat{x} + k_{y} \hat{y}\right) \quad (i = 1, 2),
\end{equation} 
is the wave-vector of the electromagnetic wave in medium 1 and 2 respectively, and $\mathbf{E}_{i}$ ($i = 1,2 $) the electric field. 
	
Using the expression (\ref{eqnki}) for wave-vector $\mathbf{k}_{i}$ and (\ref{elecfield}) for the electric field $\mathbf{E}_{i}$ in the vector equation (\ref{maxwelleqninmedium}) leads to scalar equations for each component of the electric fields ${\bf E}_1$ and ${\bf E}_2$.
Using the boundary condition $E_{1y} = E_{2y} = E_{0y}$, these equations are simplified to 
\begin{subequations}
\label{eqset1simplified}
\begin{align}
\left(k^2_{y} - \frac{\omega^2}{c^2}\epsilon_{1xx}\right)E_{1x} + \left(-i k_{1x} k_{y} - \frac{\omega^2}{c^2} \epsilon_{1xy}\right)E_{0y} = 0,\label{eqnE1xEy_1}\\
\left(-i k_{1x}k_{y} - \frac{\omega^2}{c^2}\epsilon_{1yx}\right)E_{1x} + \left(-c^2 k^2_{1x} - \frac{\omega^2}{c^2} \epsilon_{1yy}\right)E_{0y} = 0;\label{eqnE1xEy_2}
\end{align}
\end{subequations}
and
\begin{subequations}
\label{eqset2simplified}
\begin{align}
\left(c^2 k^2_{y} - \frac{\omega^2}{c^2}\epsilon_{2xx}\right)E_{2x} + \left(+i k_{2x} k_{y} - \frac{\omega^2}{c^2} \epsilon_{2xy}\right)E_{0y} = 0,\label{eqnE2xEy_1}\\
\left(+i k_{2x}k_{y} - \frac{\omega^2}{c^2}\epsilon_{2yx}\right)E_{2x} + \left(-k^2_{2x} - \frac{\omega^2}{c^2} \epsilon_{2yy}\right)E_{0y} = 0.\label{eqnE2xEy_2}
\end{align}
\end{subequations}

 Eqs.\ (\ref{eqset1simplified}) and (\ref{eqset2simplified}) will have non-trivial solutions if the determinants of the matrices of coefficients in these
equations vanish.  These requirements lead to the following expressions for $k^2_{1x}$ and $k^2_{2x}$:
\begin{subequations}
\begin{align}
k^2_{1x} &= \frac{1}{\epsilon_{1xx}}\left[k^2_{y}\epsilon_{1yy} - \frac{\omega^2}{c^2}\left(\epsilon^2_{1xx} + \epsilon^2_{1xy}\right)\right], \label{eqnk1x}\\
k^2_{2x} &= \frac{1}{\epsilon_{2xx}}\left[k^2_{y}\epsilon_{2yy} - \frac{\omega^2}{c^2}\left(\epsilon^2_{2xx} + \epsilon^2_{2xy}\right)\right]. \label{eqnk2x}
\end{align}
\end{subequations}
 
We also have the boundary conditions that the normal components of the displacement $\mathbf{D}_{x} = \left(\epsilon\mathbf{E}\right)_{x}$,  the normal $\mathbf{B}_{x}$, and the
tangential components of the magnetic field $\mathbf{B}_{z},\mathbf{B}_{y}$ must be continuous across the interface.   These conditions lead to the equations
${E}_{10} = {E}_{20}$, 
$\sum_{\alpha = x,y}\epsilon_{1x\alpha}E_{1\alpha} = \sum_{\alpha = x,y}\epsilon_{2x\alpha}E_{2\alpha}$, and
$k_{1x} E_{1y} + i k_{y} E_{1x} = -k_{2x}E_{2y} + i k_{y}E_{2x}$,
These equations, along with the requirement that the tangential component of the electric field be continuous,  give the following expressions for the ratios of the x-component to
the y component of the electric field in each component ($i = 1,2$):
\begin{subequations}
\begin{align}
\frac{E_{1x}}{E_{0y}} &= \frac{k_{y}\left(\epsilon_{1xy} - \epsilon_{2xy}\right) + i \epsilon_{2xx}\left(k_{1x} + k_{2x}\right)}{k_{y}\left(\epsilon_{2xx} - \epsilon_{1xx}\right)}, \label{ratioE1xtoEy} \\
\frac{E_{2x}}{E_{0y}} &= \frac{k_{y}\left(\epsilon_{2xy} - \epsilon_{1xy}\right) - i \left(k_{1x} + k_{2x}\right)\epsilon_{1xx}}{k_y\left(\epsilon_{1xx} - \epsilon_{2xx}\right)}. \label{ratioE2xtoEy}
\end{align}
\end{subequations}
Finally, using Eqs.\ (\ref{eqnk1x}, (\ref{eqnk2x}) and (\ref{ratioE1xtoEy}) in either (\ref{eqnE1xEy_1}) or  (\ref{eqnE1xEy_2}), {\emph or} eqs.\ (\ref{eqnk1x}), (\ref{eqnk2x}) and (\ref{ratioE2xtoEy}) in either (\ref{eqnE2xEy_1}) or (\ref{eqnE2xEy_2}),
 leads to the following equation for the surface plasmon dispersion relation:
\begin{eqnarray}
\label{eq:dispersion}
&\epsilon_{1d}\sqrt{c^2 k^2_{y} - \omega^2\epsilon_{1d}\left(1 - \epsilon^2_{1f}/\epsilon^2_{1d}\right)}\left(c^2 k^2_y - \omega^2\epsilon_{2d}\right) \nonumber \\
& + \epsilon_{2d}\sqrt{c^2 k^2_{y} - \omega^2\epsilon_{2d}\left(1 - \epsilon^2_{2f}/\epsilon^2_{2d}\right)} \left(c^2 k^2_y - \omega^2\epsilon_{1d}\right) \nonumber \\ 
&- c k_{y}\left[c^2 k^2_{y}\left(\epsilon_{1f} - \epsilon_{2f}\right) + \omega^2\left(\epsilon_{1d}\epsilon_{2f} - \epsilon_{2d}\epsilon_{1f}\right)\right)] = 0.
\end{eqnarray}
Here $\epsilon_{\mathrm{i}d} = 1 - \omega^2_{pi}/(\omega^2 - \omega^2_{Bi})$ and $\epsilon_{\mathrm{i}f} = -\omega^2_{pi}\omega_{Bi}/\left[\omega\left(\omega^2 - \omega^2_{Bi}\right)\right]$.   This equation is a generalization of eq.\ (2) of Ref.\ \cite{Yu08} and includes both bulk and the surface modes. 



Before presenting numerical results, we briefly comment on the types of materials which may be described by dielectric tensors of the form (\ref{epsilon}).  
One example would be a pair
of  Drude metals in a magnetic field\cite{AshcroftMermin76}, provided damping can be neglected.
Another example would be a pair of superconductors at frequencies below their energy gaps. The dielectric function of an isotropic superconductor at frequencies well below the BCS energy gap and at temperatures much below the transition
temperature can be approximated by the Mattis-Bardeen form\cite{mattis,tinkham}
$\epsilon(\omega) = 1 - \frac{\omega_p^2}{\omega^2}$,
where
$\omega_p^2 = \frac{\Delta\sigma_n}{4\epsilon_0\hbar}$; 
here $\Delta$ is the superconducting energy gap, $\sigma_n$ is the normal state conductivity of the superconductor and $\epsilon_0$ the permittivity of free space.
However, such frequencies lie far below the effective plasma frequency; so
this form would apply only at these low frequencies.   

\begin{figure}[h!]
\includegraphics[width=\columnwidth]{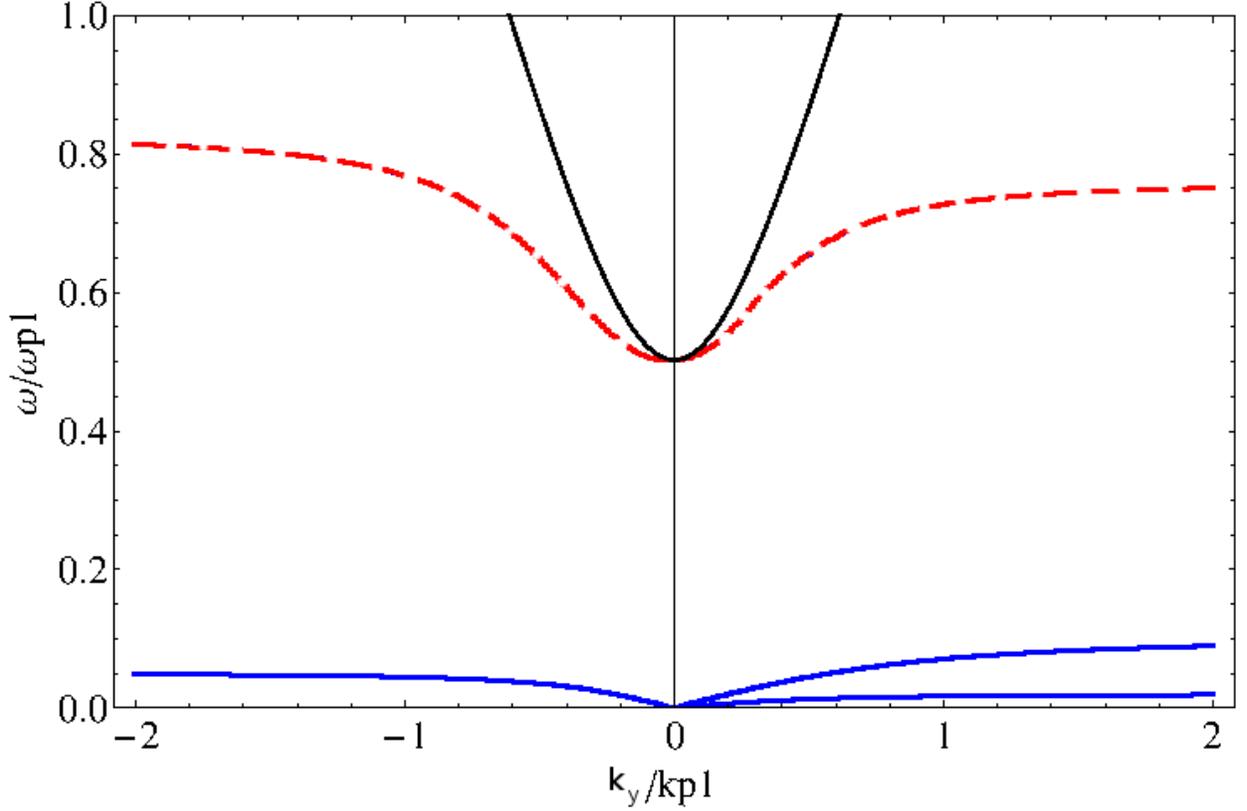}
\caption{Full curves in blue and dashed curve in red: solutions to eq.\ (12) corresponding to surface plasmon dispersion relations. 
We assume $\omega_{B1}/\omega_{p1} = \omega_{B2}/\omega_{p2} = 0.1$ and
$\omega_{p2}/\omega_{p1} = 0.5$.   Full black curve: bulk plasmon branch for metal 2.  For $0 < \omega/\omega_{p1} < 0.1$, there are two surface plasmon branches propagating to the right and only one to the left. The upper right-propagating branch, which exists
for $0.05 < \omega/\omega_{p1}< 0.1$, is one-way, propagating in the +y direction only.    
The upper surface plasmon branch exists for $\sim$ $0.59 < \omega/\omega_{p1} < 0.82$, with one-way propagation in the +y direction for
$0.79 < \omega/\omega_{p1} < 0.82$.   The horizontal axis represents the wave vector $k_y$ in units of $k_{p1} = \omega_{p1}/c$. }
\label{fig:dispersionplot1}
\end{figure}

In Fig.\  \ref{fig:dispersionplot1}, we show the various branches of the surface plasmon dispersion relation obtained from solving Eq.\ (\ref{eq:dispersion}) numerically for a particular choice of parameters: $\omega_{B1}/\omega_{p1} = \omega_{B2}/\omega_{p2} = 0.1$ and $\omega_{p2}/\omega_{p1} = 0.5$.  
 Of these surface branches, the high frequency ones  (shown in dashed red) do have a frequency range in which propagation is possible only in one direction (the negative $k_y$ direction).  However, these branches overlap with the bulk plasmon dispersion of metal 2 (shown in black) and can decay into these modes.   Because of this decay, the apparently one-way region of this dispersion relation may not be strictly one-way. 

\begin{figure}[t!]
\includegraphics[width=\columnwidth]{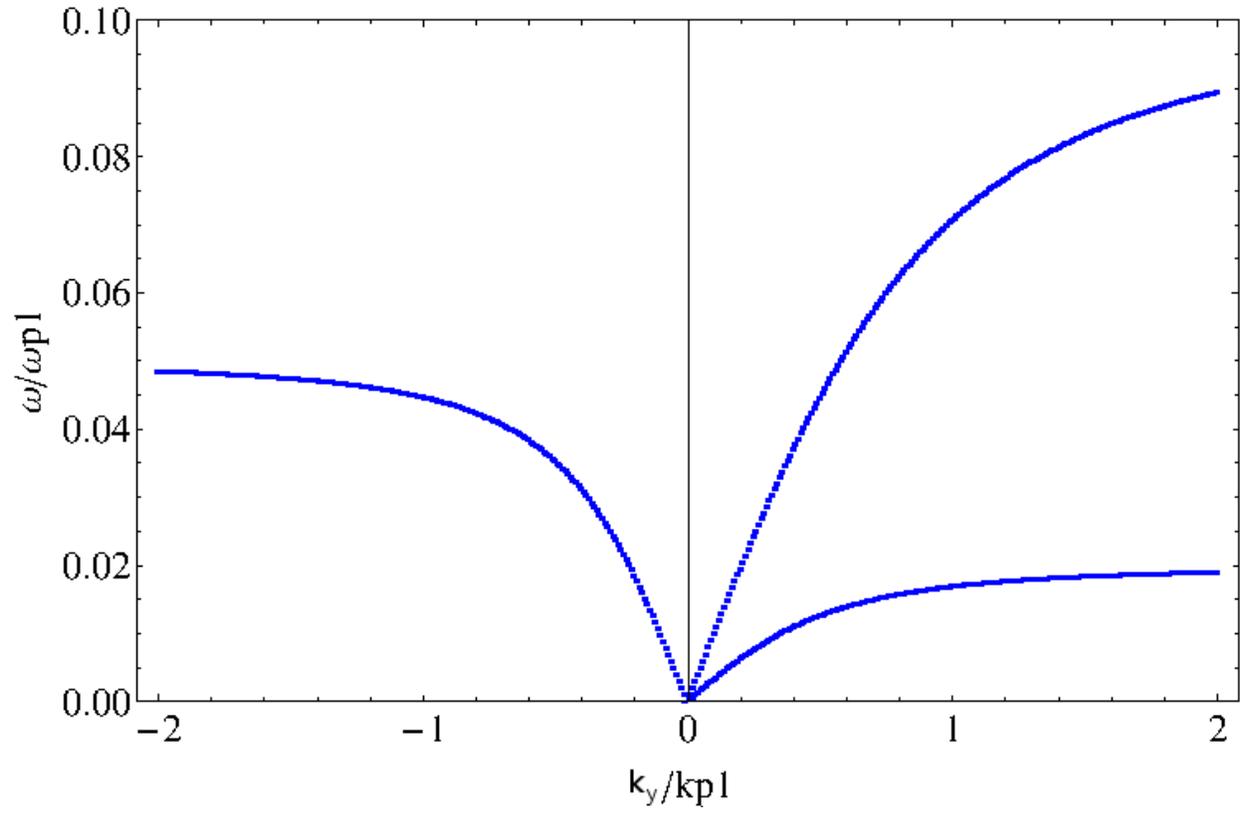}
\caption{Enlargement of the low frequency branches of the dispersion relations plotted in Fig.~\ref{fig:dispersionplot1}.} 

\label{fig:dispersionplot2}
\end{figure}

 The low frequency branches (shown in thick blue) are one way in the range $0.05 < \omega/\omega_{p1} < 0.1$.  An expansion of this frequency range is shown
in Fig.\ \ref{fig:dispersionplot2}.
In this frequency range, there exist no modes propagating to the left.  
The frequency window for the one-way propagation increases with increasing magnitude of the magnetic field, i.\ e.\ of 
the ratio $\omega_{B1}/\omega_{p1} =\omega_{B2}/\omega_{p2}$.


In the lower frequency range, which has no direct analog in the metal/vacuum system, the allowed frequency range for one-way wave propagation is $\omega_{B2} < \omega < \omega_{B1}$.
In this regime,  $\epsilon_{1d}(\omega)> 0$ and $\epsilon_{2d}(\omega) < 0$.   Thus, roughly speaking, metal 1 behaves like an (anisotropic) insulator while
metal 2 behaves like a metal.   The surface plasmon in this regime is thus propagating at the interface between a metal and a non-metal.   In the ``non-metal,'' there will still be bulk electromagnetic modes.
The surface plasmon modes in this regime can probably again decay into these bulk modes, besides experiencing any single-particle damping and radiative losses.


Thus, by generalizing the approach of Ref.\  \cite{Yu08}, we have shown that one-way electromagnetic wave propagation can occur at the interface of two dissimilar Drude metals in a magnetic field.  Such a waveguide, being based on 3D bulk materials, and involving the
interface between the two metals, should be straightforward to realize experimentally.    

In the higher-frequency range, there would probably be some radiation from these one-way modes.
But this radiation may be of little
quantitative importance in typical Drude metals, which have a finite relaxation time.   In such metals, the finite $\tau$ will lead to dissipative losses which may be more important than the radiative losses.  The modes may also be damped by surface disorder.

We have also demonstrated one-way surface modes in a low frequency regime which is not present in the metal-vacuum system.  In this low frequency regime, one metal behaves like an anisotropic insulator (with zero conductivity in the static limit), while the other
still behaves like an anisotropic metal.   This would be fascinating frequency range to study experimentally.

For typical metals with electron density $n_{ei}$ of the order $10^{28}$ m$^{-3}$ and effective mass $m_i^{*}$ of order $0.1$m$_0$, where m$_0$ is the bare electron mass, the plasma frequency $\omega_{pi} = \sqrt{n_{ei} e^2/m_i^{*}\epsilon_0}$ is of the order $10^{16}$ Hz and a ratio of $\omega_{Bi}/\omega_{pi} = 0.1$ corresponds to large B-fields of order $10^4$ T. We note that for moderate, lab values of B field of 5 - 10 T there continue to exist two distinct regimes of one-way propagation, though the low frequency branch supporting one-way propagation occurs at much lower 
frequencies.

Finally, we discuss the application of this approach to superconductors at finite magnetic field. At finite magnetic fields, one expects the dielectric function to become a tensor dielectric function of the form (2), since the assumption $\tau \rightarrow \infty$ is exactly correct.  
The magnetic field would certainly have to satisfy, at minimum, $\hbar\omega_{Bi} < \Delta_i$ in the i$^{th}$ superconductor.
If these conditions are satisfied, and if it is possible to have superconductors with two different  $\omega_{Bi}$, then one-way modes might occur in the low-frequency regime mentioned above, but with no single-particle damping.    
There would still, however, be radiative damping.   
The possibility of finding one-way modes at the interface between two superconductors is intriguing and should be explored further, as should the
unusual low-frequency non-reciprocal surface modes at this interface.



This work was supported through DOE grant No. DE-FG02-07ER46424, with additional suport from the NSF MRSEC at Ohio State (Grant
DMR-0820414).  We thank Prof. Pak Ming Hui for helpful discussions.

\end{document}